# Dynamic Taint Tracking using Partial Instrumentation for Java Applications

Manoj RameshChandra Thakur
University of California Los Angeles
manojrthakur@cs.ucla.edu

*Abstract*—**Dynamic taint tracking is the process of assigning label to variables in a program and then tracking the flow of the labels as the program executes. Dynamic taint tracking for java applications is achieved by instrumenting the application ie. adding parallel variable for each actual variable of the program and inserting additional bytecode instructions to track the flow of the parallel variables. In this paper we suggest partial instrumentation to achieve dynamic taint tracking with reasonable runtime overhead. Partial instrumentation involves instrumenting only parts of a java application, which are within the scope of a predefined source and sink set. Partial instrumentation is performed at the granularity level of a method. We use PetaBlox[2], a large-scale software analysis tool, which internally uses Datalog[3], to perform static analysis and infers all the methods within the scope of source and sink sets and a modified version of Phosphor[1] to achieve partial instrumentation. Test runs performed on some of the Dacapo benchmarks[5] show a significant performance improvement over the version of Phosphor that performs complete instrumentation.**

## I. INTRODUCTION

Dynamic taint tracking involves using additional taint information associated with program variables, methods and class objects. Even though dynamic taint tracking is an effective means of tracking the flow of tainted information through a program, it suffers from a significant runtime overhead. The runtime overhead can be as high as 62%, as is the case with the *pmd* Dacapo benchmark. Static taint analysis techniques use information gathered before runtime to understand and infer parts of the program that can cause tainted information to flow into sensitive parts of the program. These techniques however suffer from the problem of accuracy, mainly due to the complicated and dynamic nature of applications. We present a hybrid technique that uses results generated from static analysis on the java byte-code to perform partial instrumentation to enable dynamic taint tracking. The partial instrumentation adds taint tracking to only a minimal subset of the methods that will be executed by the applications at runtime. Partial instrumentation based on static analysis enables us to gain performance improvement. We use PetaBlox[2] to perform our static analysis and Phosphor[1] to partially instrument the application byte code. Figure. 1 shows the individual components of our tool and the order in which they are executed.

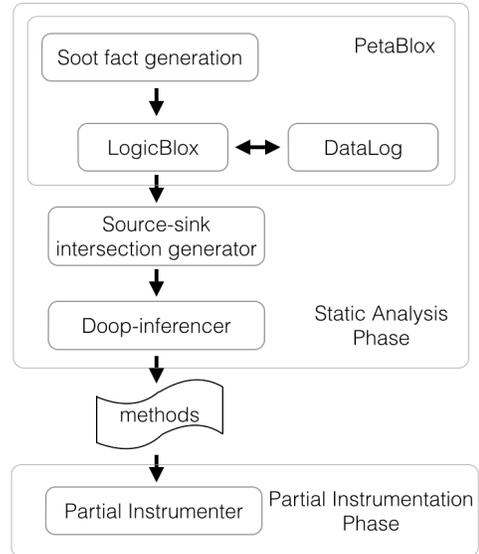

Fig. 1. Individual components of the dynamic taint analysis tool using partial instrumentation.

In the above figure, *soot-fact generation, logic-blox, datalog*, *source-sink intersection generator* and *doop-inferencer* are part of the static analysis phase. The result of the static analysis is a *methods* file that specifies all methods that are to be instrumented. The *additional-methods-inferencer* and *partial-instrumenter* are parts of the partial instrumentation phase. *Additional methods inference* identifies additional methods that need to be instrumented apart from the ones mentioned in the methods file. *Partial-instrumenter* performs partial instrumentation on the java byte code based on the methods file and output of the *additional-methods-inference* component. The remaining part of the paper is organized as follows: Section II describes components corresponding to static analysis. Section III explains the output format of the static analysis. Section IV the partial instrumentation phase. Section V explains the steps to be followed to execute the static analysis and the partial instrumentation phase. Section VI explains the experiments performed on the Dacapo benchmarks[5]. Section VII explains the limitations of our work and suggests some enhancements that we intend to implement as a part of our future work followed by section VIII which concludes our work.

## II. STATIC ANALYSIS

Static analysis is performed using the PetaBlox tool. PetaBlox at its core uses Datalog for analysis. All forms of analysis performed using PetaBlox involve writing DataLog queries to retrieve the required information about the given application. We use static analysis to determine a list of all the methods that will be executed within the scope of a source and a sink. A source is defined as a method that returns a value from an untrusted source like user input or file input stream. A sink is defined as a sensitive method that expects all its inputs to come from trusted sources and will fail otherwise. We compute the source-sink intersection based on the following tables which represent the source-seeds and sink-seeds respectively.

TABLE I
Methods used to generate source set

| Source seeds |
| --- |
| <java.io.InputStream: int read(byte[])> |
| <java.io.InputStream: int read(byte[],int,int)> |
| <java.io.BufferedReader: java.lang.String readLine()> |
| <java.io.BufferedReader: int read()> |
| <java.io.BufferedReader: int read(char[],int,int)> |
| <java.io.Reader: int read()> |
| <java.io.BufferedReader: java.lang.String readLine()> |
| <java.io.Reader: int read(char[],int,int)> |
| <java.io.Reader: int read(char[])> |
| <java.io.Reader: int read(java.nio.CharBuffer)> |

TABLE II
Methods used to generate sink set

| Sink seeds |
| --- |
| <java.io.OutputStream: void write(int)> |
| <java.io.OutputStream: void write(byte[])> |
| <java.io.OutputStream: void write(byte[],int,int)> |
| <java.io.PrintStream: void println(java.lang.String)> |
| <java.io.PrintStream: void println(java.lang.Object)> |
| <java.io.BufferedWriter: void write(char[],int,int)> |
| <java.io.BufferedWriter: void write(int)> |
| <java.io.BufferedWriter: void write(java.lang.String,int,int)> |
| <java.lang.Runtime: java.lang.Process exec(java.lang.String[])> |

We define source-set and sink-set based on the following recursive DataLog queries:

```
Source(x) <- CallGraphEdgeModified(x,y), SourceSeed(y).
Source(x) <- Source(y), CallGraphEdgeModified(y,x).

Sink(x) <- CallGraphEdgeModified(x,y), SinkSeed(y).
Sink(x) <- CallGraphEdgeModified(x,y), Sink(y).
```

In the above queries *CallGraphEdgeModified* predicate represents a *calls* relationship ie. *CallGraphEdgeModified(x,y)* implies that method *x* calls method *y*. The Sink and Source represent the sink and source set respectively. Thus source-set consists of all methods that call methods in source-seed or methods that are called from methods in source-set and sink-set consists of all methods that call methods in sink-seed or methods that call methods in sink-set. After the source and sink set is computed we find the intersection using the following query.

```
Intersection(x) <- Sink(x), Source(x).
```

The above query adds all methods that are in both source and sink sets into the Intersection set.

The individual components of the static analysis are as follows:

### A. Soot fact generation

This component loads facts about the given application into the DataLog database. Facts about the application include information like super class of a given class, methods defined by a class, parameters to a method etc. An example of facts generated by soot-fact-generation is as follows:

```
FieldModifier(?modifier, ?field) ->
        FieldSignatureRef(?field),
        ModifierRef(?modifier).
```

The above predicate `FieldModifier` holds the access modifier `?modified` for a particular field `?field` of a class.

### B. LogicBlox

LogicBlox is the DataLog engine used by PetaBlox. Analysis and the corresponding results mentioned in this paper are based on the version 3.10.21 of LogicBlox.

### C. Source-Sink Intersection Generator

This component works on a modified version of the call-graph generated by soot-fact-generation and DataLog queries. The call-graph generated by the soot-fact-generation and DataLog queries has the following format:

```
CallGraphEdge(?callerCtx, ?invocation, ?calleeCtx,
?method)
```

In the format above `callerCtx` and `calleeCtx` refer to context in which the callee is called from the caller. `Invocation` is the actual call-site where the function is called and `method` is the function that is called. The issue with this format is that we cannot apply our recursive queries to `CallGraphEdge` to generate the source and sink sets. Hence we first convert the `CallGraphEdge` to `CallGraphEdgeModified` which has the following format.

```
CallGraphEdgeModified(caller, callee)
```

In the above format `caller`, `callee` are both method signatures. We use the following queries to generate the `CallGraphEdgeModified` predicate

```
CallGraphEdgeModified(vinmethod, callee) <-
CallGraphEdge(_,vinv,_, callee),
VirtualMethodInvocation:In(vinv, vinmethod).
```

```
CallGraphEdgeModified(sinmethod, callee) <-
CallGraphEdge(_,sinv,_,callee),
StaticMethodInvocation:In(sinv, sinmethod).

CallGraphEdgeModified(spcinmethod, callee) <-
CallGraphEdge(_,spcinv,_, callee),
SpecialMethodInvocation:In(spcinv, spcinmethod).

CallGraphEdgeModified(optinmethod, callee) <-
CallGraphEdge(_,optinv, _, callee),
OptSpecialMethodInvocationBase(optinv, var),
Var:DeclaringMethod(var, optinmethod).

CallGraphEdgeModified(optinmethod, callee) <-
CallGraphEdge(_,optinv,_,callee),
OptVirtualMethodInvocationBase(optinv, var),
Var:DeclaringMethod(var, optinmethod).

CallGraphEdgeModified(caller, callee) <-
VirtualMethodInvocation(_, callee, caller).

CallGraphEdgeModified(caller, callee) <-
StaticMethodInvocation(_, callee, caller).

CallGraphEdgeModified(caller, callee) <-
SpecialMethodInvocation:In(inv, caller),
SpecialMethodInvocation:Signature(inv,callee).
```

Once the `CallGraphEdgeModified` predicate is created, we compute the source and sink sets as mentioned earlier.

### D. Doop-Inferencer

This component infers additional methods that need to be instrumented. The working of this component is influenced by the way Phosphor performs instrumentation. Doop-inference adds method in the list of methods to be instrumented (L) in the following cases:

1. A method that initializes, modifies or re-initializes a class field that is as array. The rationale behind adding such methods to L is that for each class field, phosphor adds a parallel array of taint values to store taint corresponding to each element of the array. Missing instrumentation for such methods will lead to inconsistencies in the state of the array of taint values.

2. A method that overrides a method that is already in L. Adding these methods solves the following issues:

   a. Inconsistent type hierarchy: instrumenting the super class and not instrumenting the overridden method in the subclass can lead to cases where the instrumented caller method expects to be calling an instrumented method, however the actual overridden method that will be called is not instrumented.

   b. Methods called from standard JRE code: partial instrumentation assumes that all the JRE specific code is fully instrumented. Thus if a JRE specific code calls a method on a particular JRE class, the overridden version of which is not instrumented it may lead to either *java.lang.NoSuchMethodError* or *java.lang.AbstractMethodError*.

Apart from adding the methods mentioned above we also include all JRE methods that are callees in the callgraph generated by static analysis. JRE methods include all methods that belong to a class in the package 'java.*', 'sun.*' and 'javax.*'. These methods won't be instrumented by Phosphor but will used to infer application methods that override standard JRE methods that will be executed by the program.

This is more specifically to solve the issue mentioned in 2b. The DataLog query to find all such methods is as follows:

```
CalledFromJava(x) <- CallGraphEdgeModified(y,x),
MethodSignatureRef:Value(y:ystr),(string:like(ystr,
"<java.%"); string:like(ystr, "<sun.%");
string:like(ystr,"<javax.%")).

CalledFromJava(x) <- CallGraphEdgeModified(_,x),
MethodSignatureRef:Value(x:xstr),(string:like(xstr,
"<java.%"); string:like(xstr, "<sun.%");
string:like(xstr,"<javax.%")).
```

3. A method that either returns or expects as parameters an array of dimension greater than one and is called from an instrumented method. In order to track taint for an array whose dimension is greater than one, phosphor converts the array into an array of *MultiDTaintedArray* objects. If an instrumented method is calling a method that returns or accepts as parameter a multi-dimensional array then it will expect and pass an array of *MultiDTaintedArray* objects and the method call will fail.

## III. METHODS FILE

The result of the static analysis is a *methods* file that lists all methods that need to be instrumented to achieve dynamic taint tracking. The *methods* file lists methods in the following format:

```
<com.ibm.icu.util.ULocale$IDParser: void append(char)>
```

The above line specifies that the method `append` of owner class `com.ibm.icu.util.ULocale$IDParser` which has return type of `void` and takes a `character` as parameter needs to be instrumented. During instrumentation phase the methods signature is matched with that mentioned in methods file to check if it needs to be instrumented.

## IV. PARTIAL INSTRUMENTATION

### A. Issues with partial instrumentation

Before explaining the changes required for partial instrumentation to Phosphor, we will first discuss the issue associated with partially instrumenting a java application.

1. Caller-callee contract: with respect to Phosphor there could exist two cases when instrumenting a method:

   a. A method that, as parameters, either accepts nothing or only object or object array or array with dimension greater than one. For each class that phosphor instruments, it adds a taint variable to hold the taint value of an object of that class, apart from the taint values for the member fields. Thus for this case phosphor doesn't need to add additional variables to hold the taint value of parameters. It must be noted that for arrays with dimension greater than one phosphor already creates an new single dimensional array of *MultiDTaintedArray* objects each of which store the taint of the nested arrays they encapsulate. In this case phosphor doesn't change the signature of the method after instrumentation.

   b. A method that, as parameter, accepts primitive types or primitive single dimensional arrays. In this case phosphor needs to add additional taint variables as parameter values to track taint for parameter values. Thus phosphor adds a new method with the name obtained by appending `$$INVIVO_PC` to the original method name and adding new parameters for tracking taint for each of the primitive parameter values and single dimensional arrays. Phosphor keeps a method with the original signature, which does nothing but call the instrumented version of the method with default taints.

With partial instrumentation potential mismatch in expectations can exist between the caller and callee. More specifically when the caller is instrumented (including cases where caller's signature has changed) it will always expect the callee to be instrumented and it its not we might get a *java.lang.NoSuchMethodError*. It must be noted that this problem will exist only if the signature of the callee would have changed had it been instrumented. The problem of mismatch in expectation between caller and callee will not exist if the caller is un-instrumented because irrespective of whether callee is instrumented or not the method with original signature will always exist.

2. Unboxing return value: Phosphor wraps all primitive values that are returned from methods into Objects which hold the original value along with the taint value for that variable. Consider an un-instrumented method that calls a instrumented methods that returns a *boolean* value and then uses that return value in a conditional statement. In this case the returned value will be stored as the wrapper object *TaintedBoolean* on the local variable stack. However since the method is un-instrumented it would expect a *boolean* instead of the *TaintedBoolean* on the stack and will fail.

3. Reflection: Reflection can lead to potential issues related to mismatch of expectations between caller and callee mainly because the method that is actually called is looked up dynamically based on the string value (ie. the name of the method) passed. The call sites in this case cannot be analyzed by normal means.

### B. Implementing partial instrumentation

In order to implement partial instrumentation, the instrumentation phase needs to be aware of all methods that need to be instrumented. All other modifications are centered around this list of methods to be instrumented. In order to achieve this we added *SelectiveInstrumentationManager*. This class maintains a list of all methods (in bytecode method format) that are to be instrumented. The list maintained by this class is populated from the *methods* file before the instrumentation phase begins. This list is then referred by the two main components of Phosphor to decide whether to instrument a particular method or not. *TaintTrackingClassVisitor* is the main ClassVisitor[12] that invokes all other visitors to perform instrumentation of the bytecode. The *visitMethod* function visits each method of a class and then invokes different method visitors to perform changes related to instrumentation. We refer to out list of methods to be instrumented in this function and determine when the instrumentation needs to be skipped.

Once the method instrumentation is skipped, we need to make parallel changes at the call sites to make sure that the un-instrumented version of the methods are called. This change is performed in the *visitMethodInsn*[13] method of the *TaintPassingMV*. The *visitMethodInsn* is responsible for handling method call at the call sites and taint value propagation to the methods through parameters. It is also responsible for unboxing the return values of the primitive values returned from methods called in an un-instrumented method. If the methods called at the call-site is in the list of methods to be instrumented additional bytecode instructions are inserted to pass taint values and unbox wrapped primitive values returned from methods. If however the method called is not in the list of methods to be instrumented, no additional instructions are added to pass taint values for parameters to the method as parameters.

Finally we make changes for fixing call sites when an instrumented method calls a method that is not instrumented. Phosphor handles instrumentation in case of reflection by using a central class *ReflectionMasker*. All reflective method calls are redirected through this class ie. whenever *invoke* is called on a object of *java.lang.reflect.Method* extra bytecode instructions are added to in turn call the *ReflectioMasker* to infer the new signature of the method to be called along with the appropriate taint values for the method parameters. In order to solve the problem of caller-callee expectation mismatch in case of reflection at runtime, we make the list of methods to be instrumented available at runtime as well. And all calls to the *ReflectionMasker* are redirected to the appropriate version (instrumented/un-instrumented) of the method based on our list of methods to be instrumented.

### V. EXECUTING PARTIAL INSTRUMENTATION

Since partial instrumentation involves two separate components, namely the static analysis phase and the instrumentation phase, the whole process is a bit involved and complicated. This section explains the requirements and the exact steps to be followed to achieve dynamic taint tracking using partial instrumentation.

### A. System requirements

In machine that will be used for the analysis and partial instrumentation needs to satisfy the following requirements:
1. A minimum for 4GB of RAM
2. Java (version > 1.6)
3. Git [10]
4. Vagrant [11]

### B. Tool requirements

The following tools need to be installed before starting with partial instrumentation:
1. PetaBlox[6]: used to perform static analysis during the analysis phase and generate the *methods* file. This repository also contains the required configuration files for the virtual machine that needs to be setup for the analysis.
2. Phosphor[7,8]: used to perform the actual instrumentation. This tool performs partial instrumentation based on the *methods* file generated from static analysis. The *partial-instrumentation* branch of this repository contains the changes made to the original Phosphor tool to achieve partial instrumentation.

As mentioned earlier running an application, which is instrumented by Phosphor, refers to a fully instrumented version of JRE classes. Thus before starting the analysis you need to instrument the JRE classes. It must be noted that this is just a one-time step. This fully instrumented version of JRE is reused for running all application that are instrumented by Phosphor. In order to instrument the JRE classes execute the following:

- cd $PHOSPHOR_HOME
- java -jar phosphor.jar /Library/Java/JavaVirtualMachines/jdk1.7.0_45.jdk/ Contents/Home/jre $JRE_INST_HOME

The above commands instrument the JRE classes and store them at $JRE_INST_HOME.

### C. Steps

Once the above requirements are satisfied follow the steps below:

1. cd $PETABLOX_HOME ($PETABLOX_HOME = location where PetaBlox repository is cloned)
2. vagrant up (Prepare the virtual machine)
3. vagrant ssh (Login to the virtual machine)
4. sudo apt-get update && sudo apt-get upgrade (update linux packages)
5. logout (Logout of the virtual machine)
6. vagrant provision (Re-provision the virtual machine)
7. vagrant ssh (Login to the virtual machine)
8. cd /vagrant/doop-r160113-bin (Go to the doop folder)
9. If you are trying to reproduce the Dacapo benchmark results then execute step a. else execute b.
   a. ./exec-dacapo.sh <bm>
   b. ./exec.sh <location to application jar>
10. The output of step 9 is a *methods_inst* file that is the list of all methods that are to be instrumented. Since the files in the virtual machine are synchronized with the host machine execute the following in the host machine to copy the *methods_inst* to the Phosphor directory:
    cp $PETABLOX_HOME/doop-r160113-bin/ucla-pls/methods_inst $PHOSPHOR_HOME/methods ($PHOSPHOR_HOME = the location where the Phosphor repository is cloned)
11. cd $PHOSPHOR_HOME
12. ./instrument.sh <location of un-jarred application> <location where the instrumented version of the application will be stored>
13. cd < location where the instrumented version of the application is stored >
14. ./run-instrumented.sh $JRE_INST_HOME <main-class> <parameters if required>

### VI. Benchmark Results

This sections explains the results of the experiments performed on the Dacapo benchmarks [5].

The parent machine used for these experiments was a Apple MacBook Pro (Yosemite OS version 10.10.2) with a 2.5 GHz, Core i-5 processor and 8GB of RAM. The experiments were performed on a Oracle HotSpot JVM, version 1.7.0_71. This machine hosted a virtual machine that was used for static analysis for these experiments. This VM was a 64-bit linux virtual machine with 3GB RAM and Oracle HotSpot JVM, version 1.7.0_71.

The experiment involves measuring the running time for seven out of the fourteen Dacapo benchmarks. The benchmarks were run for ten iterations and the measurement was taken on the tenth run of each benchmark. Table IV shows the running times of the seven benchmarks when no, partial and complete instrumentation was performed respectively. The table also shows the percentage increase in running time for partial and complete instrumentation over no instrumentation.

The running times measured show a significant performance improvement (as high as 34% incase of *pmd*) for partial instrumentation over complete instrumentation.

TABLE IV
Dacapo benchmark results for no, partial and complete instrumentation

| Benchmarks | No Instrumentation (msec) | Partial Instrumentation (msec) | Complete Instrumentation (msec) |
|---|---|---|---|
| avrora | 5801 | 6377 (9.93%) | 7106 (22.5%) |
| h2 | 8743 | 9321 (6.61%) | 10333 (18.18%) |
| pmd | 1957 | 2366 (20.9%) | 3183 (62.64%) |
| xalan | 1433 | 2217 (54.71%) | 2309 (61.13%) |
| batik | 1272 | 1420 (11.64%) | 1483 (16.58%) |
| sunflow | 3821 | 4455 (16.59%) | 4968 (30.02%) |
| tomcat | 2695 | 3668 (36.10%) | 4046 (50.13%) |

### VII. Limitations and future work

Some of the limitations of our work are as follows:

1. We currently perform static analysis in the virtual machine using PetaBlox. Petablox runs out of memory for application with sizes ranging from moderate to high. For example we have been unable to make the static analysis run for the *fop* benchmark as the application that the benchmark tests is big. This limitation also restricts the scope of our experiment with *tomcat, tradebeans* and *tradesoap*. These benchmarks consists of two parts the client libraries and the server. The server is sizable in terms of the number of classes and library jars. As a result we have been unable to analyze the server part of these benchmarks. The way we have performed our experiments on *tomcat* is that we have partially instrumented the client libraries (which have reasonable size) and fully instrumented the server part. This enables us to compare the results of our experiments with that of complete instrumentation.

2. The soot-fact-generation component, which is a part of the static analysis is unable to generate facts of the *lucene-core* library which is a part of the *lusearch* and *luindex* benchmarks.

Even though the results of the experiments performed on seven of the Dacapo benchmarks results seems promising we believe there is a scope for improvement in the static analysis and partial instrumentation phases. Some enhancements that we intend to experiment with and incorporate into the existing system are as follows:

1. Currently our static analysis finds all methods within the scope of the source and sinks. These methods are the ones that we instrument. However we could use Java 8 *@OSTrusted* and *@OSUnTrusted* annotations to infer all parts of the program where a value annotated as untrusted is being cast to a variable annotated as trusted. This can be achieved using a tool developed developed by the University Of Washington [9]. This tools gives a list of call sites in a program where an unsafe downcast is performed. Having found all such sites we only need to find all methods within the scope of the source and these unsafe downcast sites, since we lose the track of taint for a variable at such unsafe downcasts. We strongly believe that this enhancement will enable us to gain further performance improvement.

2. The current version of PetaBlox uses LogicBlox version 3.10.21, which takes a significant amount of time to execute DataLog[3] queries as a part of the static analysis. We intent to experiment with the latest version of LogixBlox to improve the running time and space requirements of the static analysis and be able to analyze bigger applications with PetaBlox.

## VIII. CONCLUSION

The results from the experiments performed on the Dacapo benchmarks provide supporting evidence to the idea of using partially instrumentation to achieve efficient dynamic taint tracking. Implementing partial instrumentation has its own intricacies and complications, which we have been able to successfully address. With some enhancement, the existing tool can be used to analyze industry strength codebases and detect security issues in them.

## IX. ACKNOWLEDGEMENTS

I would like to thank my advisor Prof. Jens Palsberg for his immense support, motivation and guidance throughout the course of the project. I would also like to thank Jonathan Bell and Prof. Gail Kaiser from Columbia University for their help and support in understanding Phosphor and reproducing the Dacapo benchmark results using Phosphor. Finally I would like to thank Jake Cobb from Georgia Tech for his help and support in resolving intermittent issues with PetaBlox.